\documentclass[twocolumn,showpacs, superscriptaddress]{revtex4}
\usepackage{graphicx}
\usepackage{amsmath}
\usepackage{amsmath}
\usepackage{amsfonts}
\usepackage{amssymb}
\usepackage{amsthm}

\newcommand{\Tr}{{\textrm{Tr}}}
\newcommand{\Sp}{{\textrm{Sp}}}
\newcommand{\Real}{{\textrm{Re}}}

\newtheorem{Proposition}{Proposition}
\newtheorem{Lemma}{Lemma}

\begin{document}
\title{Operational approach to the Uhlmann holonomy}

\author{Johan {\AA}berg}
\email{J.Aberg@damtp.cam.ac.uk}
\affiliation{Centre for Quantum Computation, Department of
Applied Mathematics and Theoretical Physics, University of Cambridge,
Wilberforce Road, Cambridge CB3 0WA, United Kingdom}

\author{David Kult}
\affiliation{Department of Quantum Chemistry, Uppsala University,
Box 518, Se-751 20 Uppsala, Sweden}

\author{Erik Sj\"oqvist}
\affiliation{Department of Quantum Chemistry, Uppsala University,
Box 518, Se-751 20 Uppsala, Sweden}

\author{Daniel K. L. Oi}
\affiliation{Centre for Quantum Computation, Department of
Applied Mathematics and Theoretical Physics, University of Cambridge,
Wilberforce Road, Cambridge CB3 0WA, United Kingdom}
\affiliation{SUPA, University of Strathclyde, Glasgow G4 0NG, United
Kingdom}

\date{\today}

\begin{abstract}
We suggest a physical interpretation of the Uhlmann amplitude of a
density operator.  Given this interpretation we propose an operational
approach to obtain the Uhlmann condition for parallelity. This allows
us to realize parallel transport along a sequence of density operators
by an iterative preparation procedure. At the final step the resulting
Uhlmann holonomy can be determined via interferometric measurements.
\end{abstract}


\maketitle
\section{Introduction}
If a quantum system depends on a slowly varying external parameter,
Berry \cite{berry84} showed that there is a geometric phase factor
associated to the path that an eigenvector of the corresponding Hamiltonian
traverses during the evolution.  These geometric phase factors were
later generalized by Wilczek and Zee \cite{wilczek84} to holonomies,
i.e., unitary state changes associated with the motion of a degenerate
subspace of the parameter-dependent Hamiltonian. In view of the Berry
phase and Wilczek-Zee holonomy, one may ask if a phase or a holonomy
can be associated with families of mixed states. This was answered in
the affirmative by Uhlmann \cite{Uhl} by introducing
``amplitudes" of density operators, and a condition for parallelity
of amplitudes along a family of density operators (for other approaches to geometric
phases and holonomies of mixed states and their relation to the Uhlmann approach, 
see Refs. \cite{Ell, ME2, Sjo, Pei, Tong, Cha, Sar, Slater, ME, rezakhani06}).

As mentioned above, the Berry phases and the non-Abelian holonomies
can be given a clear physical and operational interpretation in terms
of the evolution caused by adiabatically evolving quantum systems.
One may also consider the evolution as caused by a sequence of
projective measurements of observables with nondegenerate or
degenerate eigenvalues, giving rise to a Berry phase or a non-Abelian
holonomy, respectively. 
The physical interpretation of the Uhlmann amplitudes and their parallel
transport is less clear.
One interpretation of the Uhlmann amplitude 
\cite{Uhl2, Dittmann, ME,tidstrom03,rezakhani06} is that it
 corresponds to the state vector of a purification on a combined system and
ancilla.  Here we suggest another interpretation, where the amplitude
corresponds to an ``off-diagonal block'' of a density operator with
respect to two orthogonal subspaces.
In this framework, we address the question of how to obtain an explicitly operational
approach to the Uhlmann holonomy, which, to the knowledge of the authors,
 has not been previously considered \cite{comment1,sjoqvist06,uhlmann91}.

The structure of the paper is as follows. In Sec.~\ref{Uappr} we give
a brief introduction to the Uhlmann holonomy. In Sec.~\ref{interpr} we
introduce our interpretation of the Uhlmann amplitude and 
show that it is possible to determine the
amplitude using an interferometric approach. Given the interpretation
of the amplitude we consider an operational implementation of the
parallelity condition in Sec.~\ref{parallel}, and in Sec.~\ref{transp} we
use the parallelity condition to establish the parallel transport. 
In Sec.~\ref{prep} we present a technique to generate the states
 needed in the parallel transport procedure.
We
 generalize the approach to sequences of not faithful density
operators (operators not of full rank) and 
introduce a preparation procedure for
the density operators needed in the generalized case in Sec.~\ref{unfaith}. 
The paper 
ends with the conclusions in Sec.~\ref{concl}.

\section{\label{Uappr}Uhlmann holonomy}
Consider
a sequence of density operators
$\sigma_{1},\sigma_{2},\ldots,\sigma_{K}$ on a Hilbert space
$\mathcal{H}_{I}$. A sequence of amplitudes of these states are
operators $W_{1}, W_{2}, \ldots, W_{K}$ on $\mathcal{H}_{I}$, such
that $\sigma_{k}= W_{k}W_{k}^{\dagger}$.  In Uhlmann's terminology
\cite{Uhl}, a density operator $\sigma$ is \emph{faithful} if its range
$\mathcal{R}(\sigma)$ coincides with the whole Hilbert space, i.e., if
$\mathcal{R}(\sigma)=\mathcal{H}_{I}$ (Note that what we refer to as a
faithful operator is often referred to as an operator of ``full rank".) For the 
present we shall assume that all
density operators are faithful, and return to the question of
unfaithful operators in Sec.~\ref{unfaith}. Using the polar decomposition
\cite{LanTis} the amplitudes can be written $W_{k}=
\sqrt{\sigma_{k}}V_{k}$, where $V_{k}$ is unitary.  The gauge-freedom
in the Uhlmann approach is the freedom to choose the unitary operators
$V_{k}$.  For faithful density operators $\sigma_{k}$  adjacent amplitudes
 are parallel if and only if $W_{k+1}^{\dagger}W_{k}>0$.
 Given an initial amplitude
$W_{1}$ and the corresponding unitary operator $V_{1}$, the
parallelity condition uniquely determines the sequence of amplitudes
$W_{1}, W_{2},\ldots, W_{K}$, and unitaries
$V_{1},V_{2},\ldots,V_{K}$. The Uhlmann holonomy of the sequence of
density operators is defined as $U_{\textrm{Uhl}} = V_{K}V_{1}^{\dagger}$
\cite{comment2}.

The Uhlmann approach fits
 naturally within the framework of differential geometry.
The amplitudes are the elements of the total space of the fiber bundle,
  with the set of faithful density operators as the base manifold, and the set of
 unitary operators $U(N)$ as the fibers. Moreover, $WW^{\dagger}=\sigma$
 gives the projection from the total space down to the base manifold. Finally, 
given a sequence in the base manifold of density operators, the parallelity
 condition $W_{k+1}^{\dagger}W_{k}>0$ induces a unique sequence in
 the total space, leading to an element of the fiber as the resulting holonomy.

\section{\label{interpr}Interpretation of the Uhlmann amplitude}
As mentioned above, our first task is to find a physically meaningful
interpretation of the Uhlmann amplitude. We regard the density operators in the
given sequence as operators on a Hilbert space
$\mathcal{H}_{I}$ of finite dimension $N$. In addition, we
append a single qubit with Hilbert space $\mathcal{H}_{s} =
\Sp\{|0\rangle, |1\rangle\}$, with $|0\rangle$ and $|1\rangle$
orthonormal, and where $\Sp$ denotes the linear span. The total
Hilbert space we denote
$\mathcal{H}=\mathcal{H}_{I}\otimes\mathcal{H}_{s}$. Note that
$\mathcal{H}$ can be regarded as the state space of a
single particle in the two paths of a Mach-Zehnder interferometer,
where $\mathcal{H}_{I}$ corresponds to the internal degrees of freedom
(e.g., spin or polarization) of the particle and $|0\rangle$ and
$|1\rangle$ correspond to the two paths.

We let $\mathbb{Q}(\sigma^{(0)},\sigma^{(1)})$ denote the set of
density operators $\rho$ on $\mathcal{H}=
\mathcal{H}_{I}\otimes\mathcal{H}_{s}$ such that
\begin{equation}
\label{margin}
\langle 0|\rho|0\rangle  =
\frac{1}{2}\sigma^{(0)},\quad \langle 1|\rho|1\rangle  =
\frac{1}{2}\sigma^{(1)},
\end{equation}
i.e., $\mathbb{Q}(\sigma^{(0)},\sigma^{(1)})$
consists of those states that have the prescribed ``marginal states"
$\sigma_{0}$ and $\sigma_{1}$, each found with probability one
half. We span $\mathbb{Q}(\sigma^{(0)},\sigma^{(1)})$ by varying
the ``off-diagonal" operator $\langle 0|\rho|1\rangle$.
 What freedom do we have in the choice of the operator $\langle 0|\rho|1\rangle$?
This question turns out to have the following answer.

\begin{Proposition}
\label{propo1}
$\rho\in \mathbb{Q}(\sigma^{(0)},\sigma^{(1)})$ if and only if there
exists an operator $\widetilde{V}$ on $\mathcal{H}_{I}$ such that
\begin{eqnarray}
\label{repre}
\rho &=& \frac{1}{2}\sigma^{(0)}\otimes|0\rangle\langle 0| +
\frac{1}{2}\sigma^{(1)}\otimes|1\rangle\langle 1| \nonumber\\
& &+  \frac{1}{2}\sqrt{\sigma^{(0)}} \widetilde{V}
\sqrt{\sigma^{(1)}}\otimes |0\rangle\langle 1|\nonumber\\
& &  + \frac{1}{2}\sqrt{\sigma^{(1)}}\widetilde{V}^{\dagger}
\sqrt{\sigma^{(0)}}\otimes |1\rangle\langle 0|,
\end{eqnarray}
and
\begin{equation}
\label{Dcond}
\widetilde{V}\widetilde{V}^{\dagger}\leq \hat{1}_{I}.
\end{equation}
\end{Proposition}

To prove this we use the following (see Lemma 13 in Ref.~\cite{Ann}):
 Let $A$, $B$, and $C$ be operators on $\mathcal{H}_{I}$, and let $A\geq 0$, $B\geq 0$, and
\begin{equation}
\label{sdlfv}
F = A\otimes |0\rangle\langle 0| + C\otimes |0\rangle\langle 1| +
C^{\dagger}\otimes |1\rangle\langle 0| +  B\otimes |1\rangle\langle 1|.
\end{equation}
Then $F$ is positive semidefinite if and only if
\begin{equation}
\label{conditions}
P_{\mathcal{R}(A)}CP_{\mathcal{R}(B)} =C, \quad A\geq CB^{\ominus}C^{\dagger},
\end{equation}
where $P_{\mathcal{R}(A)}$ and $P_{\mathcal{R}(B)}$ denote the
projectors onto the ranges $\mathcal{R}(A)$ and $\mathcal{R}(B)$ of
$A$ and $B$, respectively. In Eq.~(\ref{conditions}) the symbol
$B^{\ominus}$ denotes the Moore-Penrose (MP) pseudo inverse
\cite{LanTis} of $B$. The reason why the MP inverse is used is to
allow us to handle those cases when $A$ and $B$ have ranges that are
proper subspaces of $\mathcal{H}_{I}$. Note that when $B$ is
invertible, the MP inverse coincides with the ordinary inverse.

To prove Proposition \ref{propo1} we first note that if $\rho$ can be
written as in Eq.~(\ref{repre}), then $\Tr(\rho)=1$ and $\rho$
satisfies Eq.~(\ref{margin}).  If we compare Eqs.~(\ref{repre}) and
(\ref{sdlfv}), we can identify $A$, $B$, and $C$, and see that they
satisfy the conditions in Eq.~(\ref{conditions}). From this follows
that $\rho$ is positive semidefinite. We can thus conclude that $\rho$
is a density operator and an element of
$\mathbb{Q}(\sigma^{(0)},\sigma^{(1)})$.

Now, we wish to show the converse, i.e., if $\rho\in
\mathbb{Q}(\sigma^{(0)},\sigma^{(1)})$ then it can be written as in
Eq.~(\ref{repre}).  By definition it follows that we can identify $A =
\sigma^{(0)}/2$ and $B = \sigma^{(1)}/2$ in Eq.~(\ref{sdlfv}).  Since
$\rho$ is positive semidefinite it follows that $C$ has to satisfy the
conditions in Eq.~(\ref{conditions}) and thus
\begin{equation}
\label{kflmbkmn}
\frac{1}{2}\sigma^{(0)} \geq 2C{\sigma^{(1)}}^{\ominus} C^{\dagger}.
\end{equation}
Define $\widetilde{V} =
2\sqrt{\sigma^{(0)}}^{\ominus}C\sqrt{\sigma^{(1)}}^{\ominus}$. From
Eq.~(\ref{kflmbkmn}) it follows that $\widetilde{V}$ satisfies
$\widetilde{V}\widetilde{V}^{\dagger}\leq 1$. Moreover,
\begin{equation}
\frac{1}{2}\sqrt{\sigma^{(0)}}\widetilde{V}\sqrt{\sigma^{(1)}} =
P_{\mathcal{R}(\sigma^{(0)})}CP_{\mathcal{R}(\sigma^{(1)})} =  C,
\end{equation}
where the last equality follows from Eq.~(\ref{conditions}). Thus we
have shown that $\rho\in \mathbb{Q}(\sigma^{(0)},\sigma^{(1)})$ if and
only if $\rho$ can be written as in Eq.~(\ref{repre}). This proves
Proposition \ref{propo1}.

Now, consider the set of density operators
$\mathbb{Q}(\sigma,\hat{1}_{I}/N)$, i.e., when one
of the marginal states is the maximally mixed state. According to
Eq.~(\ref{repre}) it follows that $\langle 0|\rho|1\rangle =
\sqrt{\sigma}\widetilde{V}/(2\sqrt{N})$.  Note that the condition in
Eq.~(\ref{Dcond}) allows us to choose $\widetilde{V}$ as an arbitrary
unitary operator, and we thus obtain
\begin{eqnarray}
\label{Uhlfall}
\rho \equiv \mathcal{D}(\sigma,W) & =&
\frac{1}{2}\sigma\otimes|0\rangle\langle 0|
 + \frac{1}{2N}\hat{1}_{I}\otimes|1\rangle\langle 1|\nonumber\\ & & +
 \frac{1}{2\sqrt{N}}W\otimes |0\rangle\langle 1| \nonumber +
 \frac{1}{2\sqrt{N}}W^{\dagger}\otimes |1\rangle\langle 0|,\\
\end{eqnarray}
where $W$ is an arbitrary Uhlmann amplitude of the density operator
$\sigma$, i.e., $\sigma = WW^{\dagger}$.  We thus have a physical 
realization of the Uhlmann amplitude as
corresponding to the off-diagonal operator $\langle 0|\rho|1\rangle $.
Note that $\mathbb{Q}(\sigma,\hat{1}_{I}/N)$ contains more
states than those corresponding to amplitudes of $\sigma$. As will be
seen later, these other states have an important role when we consider
sequences of density operators that are not faithful.

Let us note some of the differences between the above
 interpretation of the Uhlmann amplitude and the interpretation in terms
 of purifications \cite{Uhl2, Dittmann, ME,tidstrom03,rezakhani06}.
In the latter, the amplitude corresponds to a pure state on a combination
 of the system and an ancilla, such that the density operator $\sigma$ is
 retained when the ancilla is traced over. Similarly as for the purification
 interpretation, we consider here an extension to a larger Hilbert space,
 but in a different manner.
In the purification approach we extend the Hilbert space as 
$\mathcal{H}_{I}\otimes\mathcal{H}_{I}$, while in the present approach
  we extend the space as $\mathcal{H}_{I}\otimes\mathcal{H}_{s}$
 \cite{comment3}. Since $\mathcal{H}_{s}$ is two-dimensional it follows
 that $\mathcal{H}_{I}\otimes\mathcal{H}_{s}\simeq \mathcal{H}_{I}\oplus\mathcal{H}_{I}$
, i.e., the space we use to represent the state and its amplitude
 is isomorphic to an orthogonal sum of two copies of $\mathcal{H}_{I}$
 (one for each path of the interferometer).
With respect to these two subspaces the amplitude is essentially
 carried by the off-diagonal operator,
 rather than the
 whole density operator $\mathcal{D}(\sigma, W)$.  In some sense
 the amplitude describes the nature of
 the superposition between these two subspaces, or equivalently,
 the coherence of the particle between
 the two paths of the interferometer.   
One can also note from Eq.~(\ref{Uhlfall}) that the total
 state $\mathcal{D}(\sigma, W)$ is mixed rather than pure in general.
Note that this subspace approach is closely related to the
 investigations of channels and interferometry in Refs. \cite{Ann, Oi, JA, OiJA}.
Let us finally point out that the subspace approach gives a more compact 
representation than the purification approach (if $\dim(\mathcal{H}_{I})>2$
 and $\sigma$ is full rank), in the sense that in the former case a single qubit is
 added to the system, while for latter we have to add a whole copy of the original system.

\paragraph*{Determining the amplitude.}
Given a state $\rho = \mathcal{D}(\sigma,W)$ the unitary part $V$ of the
amplitude $W =\sqrt{\sigma}V$ can be experimentally determined by first applying 
onto $\rho$ the unitary operation
\begin{equation}
\label{dkfj}
U_{tot} = \hat{1}_{I}\otimes |0\rangle\langle 0| +U\otimes|1\rangle\langle 1|,
\end{equation}
where $U$ is a variable unitary operator on $\mathcal{H}_{I}$. Next,
 a Hadamard gate is applied onto $\mathcal{H}_{s}$,
followed by a measurement to determine the probability $p$ to find the
state $|0\rangle\langle 0|$. This probability turns out to be 
\begin{equation}
\label{matn}
p = \frac{1}{2} +\frac{1}{2\sqrt{N}}\Real\Tr(\sqrt{\sigma}VU^{\dagger}).
\end{equation}
By varying $U$ until $p$ is maximized we determine $U=V$ uniquely, if $\sigma$ is faithful.
 Thus, $V$ can be operationally defined
as the unitary operator giving the largest detection probability in
this setup, indirectly determining the amplitude $W =
\sqrt{\sigma}V$. 

\section{Realization of the Uhlmann holonomy}
\subsection{\label{parallel}Parallelity}
Here, we address
the question of implementing the parallelity condition between two
amplitudes.  Consider two faithful density operators $\sigma_{a}$
and $\sigma_{b}$. As mentioned above the corresponding amplitudes are
parallel if and only if $W_{b}^{\dagger}W_{a}>0$. Let
$\{|\chi_{k}\rangle\}_{k}$ be an arbitrary orthonormal basis of
$\mathcal{H}_{I}$. We denote $|\chi_{k},x\rangle = |\chi_{k}\rangle
| x\rangle$ and $P_{x} = \hat{1}_{I}\otimes|x\rangle\langle x|$ for $x=
0,1$.  Since we use the Hilbert space
$\mathcal{H}=\mathcal{H}_{I}\otimes\mathcal{H}_{s}$ to represent a
 density operator and its amplitude,  we consider
two copies of $\mathcal{H}$ in order to compare 
the amplitudes of two different density operators.  On
$\mathcal{H}\otimes\mathcal{H}$ we define the following unitary and Hermitian operator:
\begin{eqnarray}
\label{Zdef}
Z &= & \sum_{kl}  |\chi_{k},0\rangle\langle \chi_{l},1| \otimes
| \chi_{l},1\rangle\langle \chi_{k},0|  \nonumber\\
& & + \sum_{kl}  |\chi_{l},1\rangle\langle \chi_{k},0| \otimes
| \chi_{k},0\rangle\langle \chi_{l},1| \nonumber\\
& & + P_{0}\otimes P_{0} + P_{1}\otimes P_{1}.
\end{eqnarray}
For $\rho_{a}= \mathcal{D}(\sigma_{a},W_{a})$ and
$\rho_{b}= \mathcal{D}(\sigma_{b},W_{b})$,
\begin{eqnarray}
\label{Edef}
E & = & \Tr(Z\rho_{b}\otimes\rho_{a})\nonumber\\
& = & \frac{1}{2} + \frac{1}{2N}\Real\Tr(W_{b}^{\dagger}W_{a}).
\end{eqnarray}
This means that the maximal value of the real and
non-negative quantity $E$ is reached when $W_{b}$ is parallel to
$W_{a}$.

\begin{figure}
\includegraphics[width = 8.5cm]{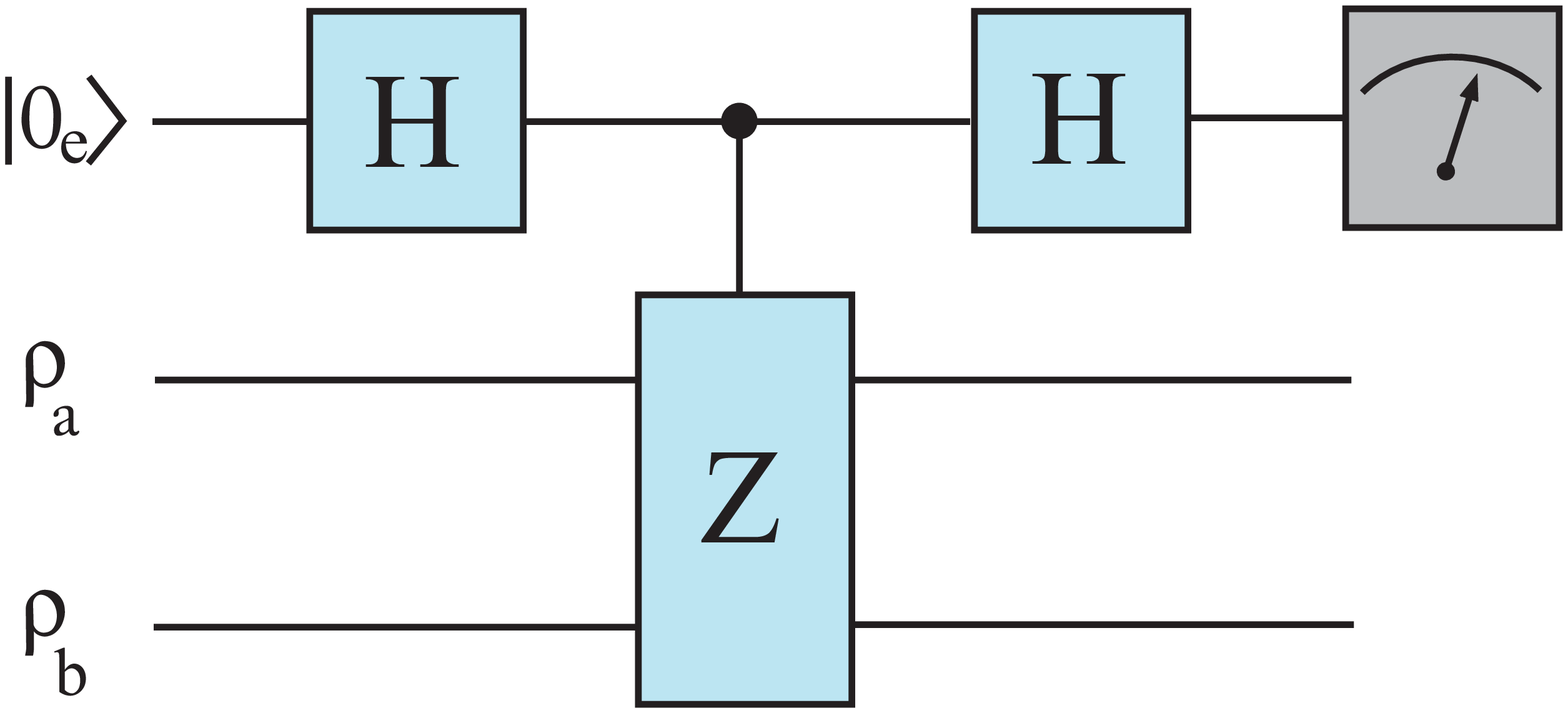}
\caption{\label{fig:figure1} (Color online) The degree of parallelity between the
amplitudes $W_{a}$ and $W_{b}$ of the states $\sigma_{a}$ and
$\sigma_{b}$, respectively, can be tested by applying this circuit
onto the states $\rho_{a} = \mathcal{D}(\sigma_{a},W_{a})$ and
$\rho_{b} = \mathcal{D}(\sigma_{b},W_{b})$ defined by Eq.~(\ref{Uhlfall}).
 A single ``extra" qubit is prepared in state $|0_{e}\rangle$ and
 a Hadamard gate is applied. Conditional on the $|0_{e}\rangle$ state
of the extra qubit, the unitary operation $Z$, defined in
Eq.~(\ref{Zdef}), is applied to $\rho_{a}\otimes\rho_{b}$. After
the application of a second Hadamard gate on the extra qubit, the
degree of parallelity between $W_{a}$ and $W_{b}$ can be inferred
from the probability to find the extra qubit in state $|0_{e}\rangle$.
Parallelity is obtained when the probability is maximal.}
\end{figure}

Now we use the fact that $Z$ is a unitary operator in order to test the 
degree of parallelity between
two amplitudes (see Fig.~\ref{fig:figure1}).  Consider an ``extra" qubit $e$ whose Hilbert space
$\mathcal{H}_{e}$ is spanned by the orthonormal basis
$\{|0_{e}\rangle, |1_{e}\rangle\}$ 
(not to be confused with
 $\mathcal{H}_{s}$ and the corresponding qubit in the construction of
 $\mathcal{D}(\sigma,W)$).
We first prepare the state $|0_{e}\rangle\langle 0_{e}|\otimes
\rho_{b}\otimes\rho_{a}$ on the total Hilbert space
$\mathcal{H}_{e}\otimes\mathcal{H}\otimes\mathcal{H} =
\mathcal{H}_{e}\otimes\mathcal{H}_{I}\otimes\mathcal{H}_{s}
\otimes \mathcal{H}_{I}\otimes\mathcal{H}_{s}$.
We apply a Hadamard gate on qubit $e$, followed by an application of
the unitary operation
\begin{equation}
U_{Z} = |0_{e}\rangle\langle 0_{e}|\otimes Z +
| 1_{e}\rangle\langle 1_{e}|\otimes\hat{1}\otimes\hat{1},
\end{equation}
i.e., an application of the unitary operation $Z$, conditioned on the
qubit $e$.  Finally, we apply the Hadamard gate on qubit $e$ and
measure the probability to find $e$ in state $|0_{e}\rangle\langle
0_{e}|$ \cite{comment4}.  This procedure results in the detection
probability
\begin{equation}
\label{detecprob}
p = \frac{1}{2}+\frac{1}{2}E= \frac{3}{4} +
\frac{1}{4N}\Real\Tr(W_{b}^{\dagger}W_{a}).
\end{equation}
Thus, the probability $p$ is maximal when $W_{b}$ is parallel to
$W_{a}$ in the Uhlmann sense.  In other
words, given the state $\rho_{a}=\mathcal{D}(\sigma_{a},W_{a})$ we
prepare various states $\rho_{b}=\mathcal{D}(\sigma_{b},W_{b})$ until
we find the amplitude $W_{b}$ that maximizes the probability $p$
\cite{comment6}. We have thus obtained an operational method to find
parallel amplitudes. One may note the similarity between this
 procedure and the method introduced in \cite{prod} to
estimate the trace of products of density operators.

The above approach is based on the fact that $Z$ is a unitary operator
and consequently corresponds to a state change. As mentioned above,
$Z$ is also Hermitian and can thus be regarded as representing an
observable.  Thus, one may consider an alternative approach where
given the state $\rho_{a}=\mathcal{D}(\sigma_{a},W_{a})$, we prepare
states $\rho_{b}= \mathcal{D}(\sigma_{b},W_{b})$ until we find the
amplitude $W_{b}$ that results in the maximal expectation value of the
observable.

It is worth to point out that the maximization which
implements the parallelity condition relates the Uhlmann holonomy with
various well known and closely related quantities.  We first note that
Eqs.~(\ref{Edef}) and (\ref{detecprob}) contain
$\Real\Tr(W_{b}^{\dagger}W_{a})$.  When this is maximized over all
amplitudes $W_{b}$ of $\sigma_{b}$ we find that 
$\sup\Real\Tr(W_{b}^{\dagger}W_{a}) = F(\sigma_{b},\sigma_{a})$, where 
$F(\sigma_{b},\sigma_{a}) = \Tr\sqrt{\sqrt{\sigma_{b}}\sigma_{a}\sqrt{\sigma_{b}}}$
is the quantum fidelity. Closely related is the transition probability
$F^{2}(\sigma_{b},\sigma_{a})$ \cite{transpr},
and the Bures metric  $d(\sigma_{a},\sigma_{b}) = \sqrt{2-2F^{2}(\sigma_{a},\sigma_{b})}$ \cite{Bures, Arkai}
that has been proved to be directly related to the Uhlmann holonomy \cite{Dittmann}.

\subsection{\label{transp}Parallel transport}
The
procedure to find parallel amplitudes allows us to obtain parallel
transport.  Suppose we are given a sequence of operators $\sigma_{j}$
on $\mathcal{H}_{I}$ for $j=1,2,\ldots, K$. We wish to construct a
sequence $\rho_{j} =\mathcal{D}(\sigma_{j},W_{j})$, such that $W_{j}$
form a parallel transported sequence of Uhlmann amplitudes. Suppose
moreover that $\rho_{1}$ is given (in order to fix the initial
amplitude $W_{1}$). We can now use the following iterative procedure:
\begin{itemize}
\item Prepare $\rho_{k}$.
\item Vary the preparations $\rho = \mathcal{D}(\sigma_{k+1},W)$ over all
amplitudes $W$ of $\sigma_{k+1}$
 until the maximum of $\Tr(Z\rho\otimes\rho_{k})$ is reached.
\item Let $\rho_{k+1} = \rho$.
\end{itemize}
After the final step $K$ we have prepared the state $\rho_{K}$ containing the amplitude $W_{K}
=\sqrt{\sigma_{K}}U_{\textrm{Uhl}}V_{1}$, where $U_{\textrm{Uhl}}$ is the Uhlmann
holonomy and $V_{1}$ is the unitary part of the chosen initial
amplitude $W_{1}$. 
The state $\rho_{K}$ can be modified by applying the unitary operator
\begin{equation}
\label{Umod}
U_{\textrm{mod}} = \hat{1}_{I}\otimes |0\rangle\langle 0| + V_{1}\otimes
| 1\rangle\langle 1|,
\end{equation}
which results in the new state
\begin{equation}
\widetilde{\rho}_{K} = U_{\textrm{mod}}\rho_{K} U_{\textrm{mod}}^{\dagger} =
\mathcal{D}(\sigma_{K},\sqrt{\sigma_{K}}U_{\textrm{Uhl}}),
\end{equation}
and hence $\langle 0|\widetilde{\rho}_{K}|1\rangle = \sqrt{\sigma_{K}}
U_{\textrm{Uhl}}/(2\sqrt{N})$.  Given this state we obtain the Uhlmann holonomy
$U_{\textrm{Uhl}}$ as the unitary operator that yields the maximal detection
probability, as described by Eq.~(\ref{matn}).

Although the iterative procedure described above is
realizable in principle, it is no doubt the case that it would be
challenging in practice, since at each step of the procedure we must
implement an optimization to find parallel amplitudes. However, the
functions we optimize over have rather favorable properties. In the case of
faithful density operators one can show that the function (taken over all
amplitudes $W_{b}$) defined by Eq.~(\ref{detecprob}) is such that there is
no local maximum except for the global maximum. In the case of not faithful
density operators the global maximum is not unique, but any of them gives the
 desired result, as shown in Sec.~\ref{unfaith}. Moreover, it still remains
the case that every local maximum is a global maximum. Hence, in both the
 faithful and unfaithful case we can apply
local optimization methods (see, e.g., \cite{Chong}).
The fact that local optimization techniques are applicable is favorable 
for practical implementations, and improve the
chances to find efficient procedures. However, a more detailed analysis would
 be required to determine what efficiency that ultimately can be obtained.
This question is, however, not considered here.

\section{\label{prep}State preparation}
\begin{figure}
\includegraphics[width = 8.5cm]{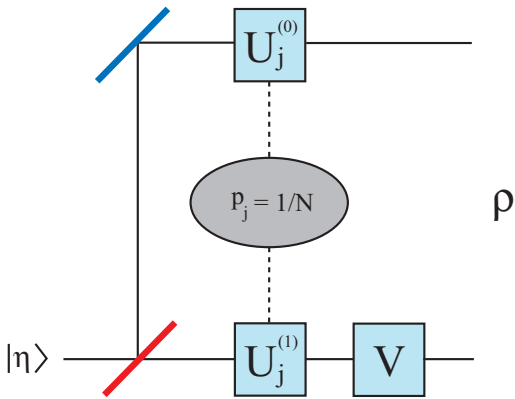}
\caption{\label{fig:figure2} (Color online) Preparation method to obtain the states
$\rho = \mathcal{D}(\sigma,W)$ that represent density operators
$\sigma$ and their amplitudes $W$, as defined in Eq.~(\ref{Uhlfall}).
The output state $\rho$ describes both the path
state and the internal state of the particle. All states
$\mathcal{D}(\sigma,W)$ can be prepared by letting a particle in a
pure internal reference state $|\eta\rangle$ and path state
$|0\rangle$ fall onto a 50-50 beam-splitter, followed by unitary
operations acting separately in the two paths on the internal state of
the particle. The application of the unitary operations have to be
coordinated by a shared output of a random generator, implementing
unitary operators $U_{j}^{(0)}$ and $U_{j}^{(1)}$ in respective path,
with probability $p_{j}=1/N$. By application of a final unitary $V$ in
path $1$ we can obtain any desired amplitude $W =\sqrt{\sigma}V$.}
\end{figure}

Since the parallel transport procedure involves repeated preparations
of states $\mathcal{D}(\sigma,W)$, with arbitrary amplitudes $W$ of
$\sigma$, we here consider preparation techniques for
such states (see Fig.~\ref{fig:figure2}). First, we show how to prepare the state $\rho =
\mathcal{D}(\sigma,\sqrt{\sigma})$. Consider the following orthogonal
but not normalized vectors:
\begin{equation}
\label{eigenv}
| \psi_{k}\rangle = \sqrt{\lambda_{k}/2}|k\rangle|0\rangle  +
| k\rangle|1\rangle/\sqrt{2N},
\end{equation}
where $\lambda_{k}$ and $|k\rangle$ are eigenvalues and corresponding
orthonormal eigenvectors of $\sigma$.  One can check that
$\sum_{k}|\psi_{k}\rangle\langle\psi_{k}| = \rho$.  The probability
distribution $(1/N,\ldots, 1/N)$ is majorized \cite{majo} by the
vector $(\lambda_{1},\ldots, \lambda_{N})$. Thus, there exists
\cite{Horn} a unitary matrix $\boldsymbol{U}$ such that
$\sum_{k}|\boldsymbol{U}_{jk}|^{2}\lambda_{k} = 1/N$ for all
$j=1,\ldots,N$ (see also Refs. \cite{nielsen00,rezakhani06}). 
Define the vectors
\begin{equation}
\label{eqeta}
| \eta_{j}\rangle = \sqrt{N}\sum_{k}\boldsymbol{U}_{jk}|\psi_{k}\rangle.
\end{equation}
One can check that these vectors are normalized. Since $\boldsymbol{U}$
is unitary it follows that $N^{-1}\sum_{j}|\eta_{j}\rangle
\langle\eta_{j}| =\rho$. Thus, $\rho$ is the result if we prepare
$|\eta_{j}\rangle$ with probability $1/N$. One can check that
$\langle \eta_{j}|P_{0}|\eta_{j}\rangle = 1/2$. Thus there  exist
normalized vectors $|\eta_{j}^{0}\rangle, |\eta_{j}^{1}\rangle \in
\mathcal{H}_{I}$, such that
\begin{equation}
\label{eqetatva}
| \eta_{j}\rangle = \frac{1}{\sqrt{2}}|\eta_{j}^{0}\rangle|0\rangle +
\frac{1}{\sqrt{2}}|\eta_{j}^{1}\rangle|1\rangle.
\end{equation}
For any normalized $|\eta\rangle\in\mathcal{H}_{I}$ there exist
unitary operators $U_{j}^{(0)}$ and $U_{j}^{(1)}$ such that
$U_{j}^{(0)}|\eta\rangle = |\eta_{j}^{0} \rangle$ and
$U_{j}^{(1)}|\eta\rangle = |\eta_{j}^{1} \rangle$.  The state $\rho$
is prepared if we apply a Hadamard gate to the state
$|\eta\rangle|0\rangle$, followed by the application of the unitary
operator $U_{j}^{(0)}\otimes |0\rangle \langle 0| + U_{j}^{(1)}\otimes
| 1\rangle \langle 1|$ with probability $p_{j}=1/N$. In terms of an
interferometric approach we thus apply a pair of unitary 
operations $U_{j}^{(0)}$, $U_{j}^{(1)}$, one in each
path of the interferometer,  with the choice of pair based
 on the output of a random generator shared between the two paths. This
procedure leads to the output density operator $\rho =
\mathcal{D}(\sigma,\sqrt{\sigma})$. To obtain a state that corresponds
to an arbitrary amplitude, i.e., $\mathcal{D}(\sigma,\sqrt{\sigma}V)$
with $V$ unitary, we just apply the unitary operation
$\hat{1}_{I}\otimes |0\rangle\langle 0| + V\otimes |1\rangle\langle
1|$ onto $\rho$. 

\section{\label{unfaith}Admissible sequences}
So far we have assumed that the density operators are faithful.  Here
we consider the generalization to admissible sequences (defined below)
of not faithful density operators \cite{Uhl}.  When the assumption of
faithfulness is removed we have to review all the steps in the
procedure.  First, we note that Eqs.~(\ref{sdlfv}) -
(\ref{Uhlfall}) are true irrespective of whether the involved density
operators are faithful or not. By using the polar decompositions
\begin{equation}
\label{dkfb}
\sqrt{\sigma_{k+1}}\sqrt{\sigma_{k}} =
\sqrt{\sqrt{\sigma_{k+1}}\sigma_{k}\sqrt{\sigma_{k+1}}}U_{k+1,k},
\end{equation}
the Uhlmann holonomy can be reformulated as $U_{\textrm{Uhl}} = U_{K,K-1}\ldots
U_{2,1}$ \cite{comment2}.  If the density operators are not faithful
then Eq.~(\ref{dkfb}) does not determine $U_{k+1,k}$
uniquely. However, if we require $U_{k+1,k}$  to be a partial isometry
 \cite{comment8} with initial space
$\mathcal{R}( \sqrt{\sigma_{k}}\sqrt{\sigma_{k+1}})$ and final space
$\mathcal{R}( \sqrt{\sigma_{k+1}}\sqrt{\sigma_{k}})$, then
Eq.~(\ref{dkfb}) uniquely determines $U_{k+1,k}$ to be the partial
isometry
\begin{equation}
U_{k+1,k}= \sqrt{\sqrt{\sigma_{k+1}} \sigma_{k}
\sqrt{\sigma_{k+1}}}^{\ominus}\sqrt{\sigma_{k+1}}\sqrt{\sigma_{k}}.
\end{equation}
If the sequence of density operators is such that the final space of
$U_{k+1,k}$ matches the initial space of $U_{k+2,k+1}$, then we may
define the Uhlmann holonomy as the partial isometry $U_{\textrm{Uhl}} =
U_{K,K-1}\ldots U_{2,1}$ \cite{Uhl}. A sequence of density operators
that results in such matched initial and final spaces constitutes an
``admissible ordered set" of density operators \cite{Uhl}. Another way
to express the condition for an admissible sequence is
\begin{eqnarray}
\mathcal{R}(\sqrt{\sigma_{k+1}}\sqrt{\sigma_{k}}) & = &
\mathcal{R}\big((\sqrt{\sigma_{k+2}}\sqrt{\sigma_{k+1}})^{\dagger}\big)
\nonumber\\
 &\equiv & \mathcal{R}(\sqrt{\sigma_{k+1}}\sqrt{\sigma_{k+2}}),
\end{eqnarray}
for $k=1,\ldots, K-2$.

Now we introduce some terminology and notation.  We say that an
operator $\widetilde{W}$ on $\mathcal{H}$ is a \emph{subamplitude} of
$\sigma$ if $\widetilde{W}\widetilde{W}^{\dagger}\leq \sigma$. It can
be shown that $\widetilde{W}$ is a subamplitude if and only if it can
be written $\widetilde{W} = \sqrt{\sigma}\widetilde{V}$, where
$\widetilde{V}\widetilde{V}^{\dagger}\leq \hat{1}_{I}$. One may note
that the physical interpretation we have constructed encompasses these
subamplitudes.  Given a density operator $\sigma$ and one of its
subamplitudes $\widetilde{W}$, we let
$\mathcal{D}(\sigma,\widetilde{W})$ denote the density operator in
Eq.~(\ref{Uhlfall}) with the amplitude $W$ replaced by the
subamplitude $\widetilde{W}$. One can see that when $\widetilde{W}$ is
varied over all subamplitudes, then
$\mathcal{D}(\sigma,\widetilde{W})$ spans all of
$\mathbb{Q}(\sigma,\hat{1}_{I}/N)$.

The following modified procedure results in the Uhlmann holonomy for
an arbitrary admissible sequence of density operators. Let
$\sigma_{1},\ldots, \sigma_{K}$ be an admissible ordered sequence of
density operators.  Assume $\rho_{1} = \mathcal{D}(\sigma_{1},
\sqrt{\sigma_{1}}\widetilde{V}_{1})$ is given, where we assume that
$\sqrt{\sigma_{1}}\widetilde{V}_{1}$ is an amplitude (not a subamplitude). 
For $k
=1,\ldots,K-1$:
\begin{itemize}
\item Prepare $\rho_{k} = \mathcal{D}(\sigma_{k},\sqrt{\sigma_{k}}
\widetilde{V}_{k})$.
\item Vary the preparation of $\rho = \mathcal{D}(\sigma_{k+1},
\sqrt{\sigma_{k+1}}\widetilde{V})$ with
$\widetilde{V}\widetilde{V}^{\dagger}\leq \hat{1}_{I}$ until the
maximum of $\Tr(Z\rho\otimes\rho_{k})$ is reached.
\item Let $\widetilde{V}_{k+1} = P_{\mathcal{R}(\sqrt{\sigma_{k+1}}
\sqrt{\sigma_{k}})}\widetilde{V}$.
\end{itemize}
After the final step
\begin{equation}
U_{\textrm{Uhl}} = \widetilde{V}_{K}\widetilde{V}_{1}^{\dagger}.
\end{equation}
Note that we may reformulate the second step as a variation of $\rho$
over all $\mathbb{Q}(\sigma_{k+1},\hat{1}_{I}/N)$, and thus we vary
over all possible subamplitudes of $\sigma_{k+1}$.  Note also that, by
the very nature of the problem, the sequence of density operators
$\sigma_{1},\ldots,\sigma_{K}$ is known to us. Thus, the projectors
$P_{\mathcal{R}(\sqrt{\sigma_{k+1}}\sqrt{\sigma_{k}})}$, that we are
supposed to apply in each step of the preparation procedure, are also
known to us. After the last step we ``extract" the Uhlmann holonomy
as described below.

To outline of the proof of the modified procedure we first note the
following fact.
\begin{Lemma}
\label{fix}
Let $A$ be an arbitrary operator on $\mathcal{H}_{I}$. If
$\widetilde{V}$ is such that it maximizes
$\Real\Tr(A\widetilde{V}^{\dagger})$ among all operators on
$\mathcal{H}_{I}$ that satisfies
$\widetilde{V}\widetilde{V}^{\dagger}\leq\hat{1}_{I}$, then
\begin{equation}
\widetilde{V} =  \sqrt{AA^{\dagger}}^{\ominus}A + Q,
\end{equation}
where $Q$ satisfies
$P_{\mathcal{R}(A)}^{\perp}QP_{\mathcal{R}(A^{\dagger})}^{\perp} = Q$,
and where $P_{\mathcal{R}(A)}^{\perp}$ denotes the projector onto the
orthogonal complement of the range $\mathcal{R}(A)$ of $A$.
\end{Lemma}
The following lemma is convenient for the proof of the modified procedure.
\begin{Lemma}
\label{iter}
Let $\sigma_{1},\sigma_{2},\ldots,\sigma_{K}$ be an admissible
sequence of density operators, and suppose that the operator
$\widetilde{V}_{k}$ satisfies
\begin{equation}
\label{indhyp}
\widetilde{V}_{k}\widetilde{V}_{k}^{\dagger} =
P_{\mathcal{R}(\sqrt{\sigma_{k}}\sqrt{\sigma_{k-1}})}.
\end{equation}
It follows that if $\widetilde{V}$ maximizes
$\Real\Tr(\sqrt{\sigma_{k+1}}\sqrt{\sigma_{k}}\widetilde{V}_{k}
\widetilde{V}^{\dagger})$ among all
$\widetilde{V}\widetilde{V}^{\dagger}\leq \hat{1}_{I}$, then
\begin{equation}
\label{projec}
\widetilde{V}_{k+1} \equiv P_{\mathcal{R}(\sqrt{\sigma_{k+1}}
\sqrt{\sigma_{k}})}\widetilde{V}
 = U_{k+1,k}\widetilde{V}_{k}
\end{equation}
is uniquely determined and satisfies
\begin{equation}
\label{nvsdkjvn}
\widetilde{V}_{k+1}\widetilde{V}_{k+1}^{\dagger}  =
P_{\mathcal{R}(\sqrt{\sigma_{k+1}}\sqrt{\sigma_{k}})}.
\end{equation}
\end{Lemma}

\begin{proof}
We have to prove that if there exists an operator $\widetilde{V}$ that
maximizes $\Real\Tr(\sqrt{\sigma_{k+1}}\sqrt{\sigma_{k}}
\widetilde{V}_{k}\widetilde{V}^{\dagger})$
and is such that $\widetilde{V}\widetilde{V}^{\dagger}\leq \hat{1}_{I}$,
then this operator satisfies Eq.~(\ref{projec}).  According to Lemma
\ref{fix} (with $A =
\sqrt{\sigma_{k+1}}\sqrt{\sigma_{k}}\widetilde{V}_{k}$) it follows
that we can write
\begin{equation}
\label{sdfhsh}
\widetilde{V} =
\sqrt{\sqrt{\sigma_{k+1}}\sqrt{\sigma_{k}}\widetilde{V}_{k}\widetilde{V}_{k}^{\dagger}
\sqrt{\sigma_{k}}\sqrt{\sigma_{k+1}}}^{\ominus}\sqrt{\sigma_{k+1}}\sqrt{\sigma_{k}}\widetilde{V}_{k}
+ Q,
\end{equation}
where
\begin{equation}
\label{sfbsggns}
\quad
P^{\perp}_{\mathcal{R}(\sqrt{\sigma_{k+1}}\sqrt{\sigma_{k}}\widetilde{V}_{k})}
QP^{\perp}_{\mathcal{R}(\widetilde{V}_{k}^{\dagger}\sqrt{\sigma_{k}}\sqrt{\sigma_{k+1}})}
= Q.
\end{equation}
If we combine Eq.~(\ref{indhyp}) with the assumption that the sequence
is admissible, and thus
$P_{\mathcal{R}(\sqrt{\sigma_{k}}\sqrt{\sigma_{k-1}})} =
P_{\mathcal{R}(\sqrt{\sigma_{k}}\sqrt{\sigma_{k+1}})}$, it can be
shown that
\begin{equation}
\label{srtu}
 \sqrt{\sigma_{k+1}}\sqrt{\sigma_{k}}\widetilde{V}_{k}\widetilde{V}_{k}^{\dagger}
 \sqrt{\sigma_{k}}\sqrt{\sigma_{k+1}} =
 \sqrt{\sigma_{k+1}}\sigma_{k}\sqrt{\sigma_{k+1}}.
\end{equation}
If Eq.~(\ref{srtu}) is inserted into Eq.~(\ref{sdfhsh}) we obtain
$\widetilde{V} = U_{k+1,k}\widetilde{V}_{k} + Q$.  By the properties
of the operator $Q$ in Eq.~(\ref{sfbsggns}), together with
$\mathcal{R}( \sqrt{\sigma_{k+1}}\sqrt{\sigma_{k}}\widetilde{V}_{k}) =
\mathcal{R}(\sqrt{\sigma_{k+1}}\sqrt{\sigma_{k}})$, it follows that
$\widetilde{V}$ satisfies Eq.~(\ref{projec}).

Now we have to prove that $\widetilde{V}_{k+1}$ satisfies
Eq.~(\ref{nvsdkjvn}). We again make use of the assumption that the
sequence is admissible, and we find that
\begin{eqnarray}
\label{oypkmmf}
\widetilde{V}_{k+1}\widetilde{V}_{k+1}^{\dagger} & = &
  U_{k+1,k}P_{\mathcal{R}(\sqrt{\sigma_{k}}\sqrt{\sigma_{k-1}})}U_{k+1,k}^{\dagger}\nonumber\\
  & = &
  U_{k+1,k}P_{\mathcal{R}(\sqrt{\sigma_{k}}\sqrt{\sigma_{k+1}})}U_{k+1,k}^{\dagger}\nonumber\\
  & = & P_{\mathcal{R}(\sqrt{\sigma_{k+1}}\sqrt{\sigma_{k}})},
\end{eqnarray}
where in the last equality we have used that the initial space of
$U_{k+1,k}$ is $\mathcal{R}(\sqrt{\sigma_{k}}\sqrt{\sigma_{k+1}})$.
Note that Eq.~(\ref{oypkmmf}) implies that
$\widetilde{V}_{k+1}\widetilde{V}_{k+1}^{\dagger} \leq \hat{1}_{I}$.  We
have now proved that if there exists a maximizing operator
$\widetilde{V}$, then this operator satisfies Eqs.~(\ref{projec}) and
(\ref{nvsdkjvn}). We finally have to prove that there actually does
exist such an operator.  If we let $\widetilde{V} =
U_{k+1,k}\widetilde{V}_{k}$, then the assumption of admissible
sequences can be used to show that
\begin{eqnarray}
\Real\Tr(\sqrt{\sigma_{k+1}}\sqrt{\sigma_{k}}\widetilde{V}_{k}\widetilde{V}^{\dagger})
& = &  \Real\Tr(\sqrt{\sigma_{k+1}}\sqrt{\sigma_{k}}U_{k+1,k}^{\dagger})
\nonumber\\
& = &\Tr\sqrt{\sqrt{\sigma_{k+1}}\sigma_{k}\sqrt{\sigma_{k+1}}},
\end{eqnarray}
which is the maximal value of
 $\Real\Tr(\sqrt{\sigma_{k+1}}\sqrt{\sigma_{k}}\widetilde{V}_{k}\widetilde{V}^{\dagger})$
 under the assumption that $\widetilde{V}\widetilde{V}^{\dagger}\leq
 \hat{1}_{I}$.  This proves the lemma.
\end{proof}

Lemma \ref{iter} can be used to prove the modified procedure in an
iterative manner.  We begin with the first step of the procedure.
Thus, we are given the state
$\rho_{1}= \mathcal{D}(\sigma_{1},\sqrt{\sigma_{1}}\widetilde{V}_{1})$, where
$\sqrt{\sigma_{1}}\widetilde{V}_{1}$ is an amplitude of $\sigma_{1}$,
and hence $\widetilde{V}_{1} \widetilde{V}_{1}^{\dagger} =
P_{\mathcal{R}(\sigma_{1})}$.  If we now let $\rho = \mathcal{D}(\sigma_{2},W)$, and vary $W =
\sqrt{\sigma_{2}}\widetilde{V}$ over all subamplitudes of $\sigma_{2}$,
we find the maximum of $\Tr(Z\rho\otimes\rho_{1})$ to be obtained
when we reach the maximum of
\begin{equation}
\Real\Tr(W^{\dagger}W_{1}) =
\Real\Tr(\sqrt{\sigma_{2}}\sqrt{\sigma_{1}}\widetilde{V}_{1}\widetilde{V}^{\dagger}).
\end{equation}
According to Lemma \ref{iter}, every maximizing $\widetilde{V}$ is
 such that
 $P_{\mathcal{R}(\sqrt{\sigma_{2}}\sqrt{\sigma_{1}})}\widetilde{V} =
 U_{2,1}\widetilde{V}_{1}$. Now we let $\widetilde{V}_{2} =
 P_{\mathcal{R}(\sqrt{\sigma_{2}}\sqrt{\sigma_{1}})}\widetilde{V}$ and
 prepare the state
 $\rho_{2} = \mathcal{D}(\sigma_{2},\sqrt{\sigma_{2}}\widetilde{V}_{2})$. We can
 repeat the above procedure in an iterative manner to find that
 \begin{equation}
\widetilde{V}_{K} = U_{K,K-1}\ldots U_{2,1}\widetilde{V}_{1} =
U_{\textrm{Uhl}}\widetilde{V}_{1}.
\end{equation}
Note that $\widetilde{V}_{1}$ is a partial isometry and thus may be
completed to a unitary operator $\overline{V}_{1}$ \cite{comment9}. If we apply
$U_{\textrm{mod}}$ in Eq.~(\ref{Umod}), but with $V_{1}$ replaced by
$\overline{V}_{1}$, the resulting state is
$\mathcal{D}(\sigma_{K},\sqrt{\sigma_{K}}U_{\textrm{Uhl}})$. (Note that
$\overline{V}_{1}$ is not unique, but this does not matter since
$\widetilde{V}_{1}\overline{V}_{1}^{\dagger} =
\widetilde{V}_{1}\widetilde{V}_{1}^{\dagger}$ for all such
extensions.) Now we shall extract the Uhlmann holonomy from the
state $\mathcal{D}(\sigma_{K},\sqrt{\sigma_{K}}U_{\textrm{Uhl}})$. Note
 that $U_{\textrm{Uhl}}$ is a partial isometry and that
$\sqrt{\sigma_{K}}U_{\textrm{Uhl}}$ in general is a subamplitude of
$\sigma_{K}$. In order to make sure that we indeed extract the Uhlmann
holonomy when we apply the procedure described in Sec.~\ref{interpr},
we first apply the projection
$P_{\mathcal{R}(\sqrt{\sigma_{K}}\sqrt{\sigma_{K-1}})}\otimes
| 0\rangle\langle 0| + \hat{1}_{I}\otimes |1\rangle\langle 1|$ onto the
state $\mathcal{D}(\sigma_{K},\sqrt{\sigma_{K}}U_{\textrm{Uhl}})$. By this
filtering (post selection) we obtain a new normalized state
$\overline{\rho}$, for which
\begin{eqnarray*}
\langle 0|\overline{\rho}|1\rangle & = &\mathcal{N}
 P_{\mathcal{R}(\sqrt{\sigma_{K}}\sqrt{\sigma_{K-1}})}\sqrt{\sigma_{K}}U_{\textrm{Uhl}}\nonumber\\
 & = & \mathcal{N} P_{\mathcal{R}(\sqrt{\sigma_{K}}
 \sqrt{\sigma_{K-1}})}\sqrt{\sigma_{K}}P_{\mathcal{R}(\sqrt{\sigma_{K}}\sqrt{\sigma_{K-1}})}U_{\textrm{Uhl}},
\end{eqnarray*}
where the constant $\mathcal{N}$ is a real nonnegative number, and
where the last equality follows since
$\mathcal{R}(\sqrt{\sigma_{K}}\sqrt{\sigma_{K-1}})$ is the final space
of $U_{\textrm{Uhl}}$.  If we apply the extraction procedure described in
Sec.~\ref{interpr} we find that the unitary operator $U$ that gives
the maximal detection probability is not uniquely determined.
However, by using Lemma \ref{fix} one can show that every maximizing
unitary operator $U$ can be written $U = U_{\textrm{Uhl}} + Q$, where
$P_{\mathcal{R}(\sqrt{ \sigma_{K}}\sqrt{\sigma_{K-1}})}^{\perp}
QP_{\mathcal{R}(\sqrt{\sigma_{K}}\sqrt{\sigma_{K-1}})}^{\perp} =
Q$. Hence, $P_{\mathcal{R}(\sqrt{\sigma_{K}}\sqrt{\sigma_{K-1}})}U =
U_{\textrm{Uhl}}$ is uniquely defined by this procedure. We have thus found a
modified procedure to obtain the Uhlmann holonomy for admissible
sequences of density operators.

\paragraph*{Preparation procedures for unfaithful density operators.}
As a final note concerning the generalization to unfaithful density
 operators we show that the preparation procedure described in
 Eqs.~(\ref{eqeta}) and (\ref{eqetatva}) to obtain the states
 $\mathcal{D}(\sigma,W)$ with $W$ an amplitude of $\sigma$, can be
 modified to obtain states $\mathcal{D}(\sigma,\widetilde{W})$, with
 $\widetilde{W}$ an arbitrary subamplitude of $\sigma$.  All
 subamplitudes $\widetilde{W} = \sqrt{\sigma}\widetilde{V}$ can be
 reached via $\widetilde{V}$ such that
 $\widetilde{V}\widetilde{V}^{\dagger}\leq\hat{1}_{I}$.  The set of
 operators $\widetilde{V}$ on $\mathcal{H}_{I}$ such that
 $\widetilde{V}\widetilde{V}^{\dagger}\leq\hat{1}_{I}$, forms a convex
 set whose extreme points are the unitary operators on
 $\mathcal{H}_{I}$, which follows from Lemma 21 in Ref.~\cite{Ann}.
 Thus, for every choice of $\widetilde{V}$ there exist probabilities
 $\mu_{n}$ and unitaries $V_{n}$, such that $\widetilde{V} =
 \sum_{n}\mu_{n}V_{n}$.  Hence, instead of applying the unitary
 operator $\hat{1}_{I}\otimes |0\rangle\langle 0| + V\otimes
 |1\rangle\langle 1|$ at the end of the preparation procedure, we can
 instead apply $\hat{1}_{I}\otimes |0\rangle\langle 0| + V_{n}\otimes
 |1\rangle\langle 1|$ with probability $\mu_{n}$.  This modified
 procedure results in the desired state $\rho =
 \mathcal{D}(\sigma,\widetilde{W})$.

\section{\label{concl}Conclusions}
In conclusion, we present an interpretation of the Uhlmann amplitude
that gives it a clear physical meaning and makes it a measurable
object. In contrast to previous approaches where the amplitude resides in 
the total pure state of a twofold copy of the original system (a purification),
 we suggest an alternative where the amplitude is represented by 
the coherences of a mixed state on a composite system.
This gives a more
compact representation and also allows for a direct interferometric
determination of the Uhlmann parallelity condition.
 Based on this, we reformulate the parallelity condition entirely in operational terms, which
enables an implementation of parallel transport of amplitudes along a
sequence of density operators through an iterative procedure.  At the
end of this transport process the Uhlmann holonomy can be identified
as a unitary mapping that gives the maximal detection probability in
an interference experiment.

In this paper, we consider the Uhlmann holonomy concomitant to
sequences of density operators, i.e., discrete families of density
operators. However, the Uhlmann holonomy can also be associated to
a smooth path of density operators \cite{Uhl}, e.g., the time evolution of a quantum
system. The parallel transport procedure discussed here is by its
very nature iterative, but we can form successively refined discrete
 approximations of the desired path and obtain the Uhlmann holonomy
 within any non-zero error bound.
The question is whether it is possible to find an operational parallel
 transport procedure formulated
\emph{explicitly} for smoothly parameterized families of density
operators.  We hope that the framework we suggest in this paper may serve as a starting
 point for such an attempt. Given a
family of density operators $\sigma(s)$, one could consider a differential equation
 for the evolution of
$\mathcal{D}\boldsymbol{(}\sigma(s), W(s)\boldsymbol{)}$ (defined in Eq.~(\ref{Uhlfall})) such
that $W(s)$ becomes the parallel transported amplitudes of
$\rho(s)$. However, it is far from clear whether such a differential equation could be
given a reasonable physical and operational interpretation.

\acknowledgments
J.{\AA}. wishes to thank the Swedish Research Council for financial
support and the Centre for Quantum Computation at DAMTP, Cambridge,
for hospitality.  E.S. acknowledges financial support from the Swedish
Research Council.  D.K.L.O. acknowledges the support of the
Cambridge-MIT Institute Quantum Information Initiative, EU grants RESQ
(IST-2001-37559) and TOPQIP (IST-2001-39215), EPSRC QIP IRC (UK), and
Sidney Sussex College, Cambridge.


\begin{thebibliography}{99}
\bibitem{comment1} One may note that for special admissible sequences
 of density operators, namely those where the density operators are proportional 
to projectors of fixed rank, a procedure presented in Ref.~\cite{sjoqvist06} does give
 rise to the Uhlmann holonomy. (For a more elaborate discussion see Sec.~II
 in Ref. \cite{sjoqvist06}.) However, we would like to point out that, as far as
 we can see, the  procedure described in Ref. \cite{sjoqvist06} cannot be used
 to determine the Uhlmann holonomy
for more general sequences of density operators.
Note also that an approach to obtain the phase of the Hilbert-Schmidt
 product between the initial amplitude and the parallel transported
 final amplitude, i.e., the trace of the Uhlmann holonomy invariant
 \cite{uhlmann91}, has been considered in Refs.~\cite{ME,tidstrom03}.
\bibitem{comment2} Note that in Ref.~\cite{Uhl} the Uhlmann holonomy
is defined as $\overline{U}_{\textrm{Uhl}} = V_{1}V_{K}^{\dagger}$. Thus,
$\overline{U}_{\textrm{Uhl}} = U_{\textrm{Uhl}}^{\dagger}$. Similarly, in \cite{Uhl} the
relative amplitudes are defined as $\overline{U}_{k,k+1} =
\sqrt{\sqrt{\sigma_{k}}\sigma_{k+1}\sqrt{\sigma_{k}}}^{-1}\sqrt{\sigma_{k}}\sqrt{\sigma_{k+1}}$,
and thus $\overline{U}_{\textrm{Uhl}} =
\overline{U}_{1,2}\overline{U}_{2,3}\ldots \overline{U}_{K-1,K}$.
Here, however, the relative amplitudes are defined as $U_{k+1,k}=
\sqrt{\sqrt{\sigma_{k+1}}\sigma_{k}\sqrt{\sigma_{k+1}}}^{-1}\sqrt{\sigma_{k+1}}\sqrt{\sigma_{k}}$,
and consequently $U_{\textrm{Uhl}} = U_{K,K-1}\ldots U_{3,2}U_{2,1}$.
\bibitem{comment3} For the purification approach the dimension of the ancillary space does
 not necessarily have to be the same as that of $\mathcal{H}_{I}$, although it at
 least have to be the rank of $\sigma$. Hence, 
if $\sigma$ is faithful the dimension of the ancillary space has to be at least $\dim(\mathcal{H}_{I})$.
 An analogous comment can be made concerning
 the ``subspace approach" we use here, although we have chosen equal
 dimensions since this enables an interpretation in terms of interferometers.
\bibitem{comment4} One could consider to add a variable phase shifter
to this procedure as to make it a proper interferometer. However,
this would be superfluous since $E= \Tr(Z\rho_{a}\otimes\rho_{b})$
is real and nonnegative.  This implies that there would not be any
shift of the interference fringes and hence no phase information to
obtain.
\bibitem{comment6} The parallel amplitude still gives the maximum
even if we extend the maximization to whole $\mathbb{Q}(\sigma_{b},
\hat{1}_{I}/N)$.
\bibitem{comment8} An operator $Q$ on a Hilbert
space is a partial isometry if and only if $QQ^{\dagger}$ and
$Q^{\dagger}Q$ are projectors. $Q^{\dagger}Q$
projects onto the ``initial space", which is mapped by $Q$ to the
``final space" onto which $QQ^{\dagger}$ projects.  See
Ref.~\cite{LanTis} for further details.
\bibitem{comment9} If the operator $\widetilde{V}_{1}$ on 
$\mathcal{H}$ is a partial isometry with initial space $\mathcal{I}$
 and final space $\mathcal{F}$, it can be ``completed" into a
 unitary operator by adding to it any partial isometry with initial
 space $\mathcal{I}^{\perp}$ and final space $\mathcal{F}^{\perp}$.
\end{thebibliography}
\end{document}